\documentclass[preprint,authoryear,12pt]{elsarticle}

\usepackage{amssymb}
\usepackage{amsthm,amsmath}
\usepackage{subfigure}
\usepackage[center]{titlesec}
\usepackage[margin=1in]{geometry}
\begin{document}

\section{Introduction}\label{intro}
Is environmental policy beneficial to the environment without hurting the economy? The answer is not clear-cut.
One debate is on the validity of the so-called double dividend hypothesis, this is, the claim that environmental tax could simultaneously improve the environment (the first dividend) and reduce the economic costs of the tax system (the second dividend), thereby avoiding the advocated environment-growth trade-off \citep{Parry2000}. The first dividend is commonly acknowledged. But for the second dividend, it is not guaranteed to emerge, although possible by returning environmental tax revenues through cuts in pre-existing distortionary taxes \citep{Bosquet2000,Patuelli2005}. One important issue often ignored in these studies to date is the link between pollution, human health and labor productivity. For instance, 55 million workers out of 148 million, aged 19-64, had an inability to concentrate at work because of their own illness or that of a family member in the US in 2003
\citep{Davis2005}.

Therefore, this paper focuses on the second dividend issue in an overlapping generations economy with the detrimental effect of pollution on health. Taking the environmental tax rate as given, the tax revenues recycling is interpreted as a different form, i.e., the returning towards labor income as well as pollution-mitigation activities. The second dividend is defined on per-worker output and lifetime welfare. In this context, how to allocate the environmental tax revenues to maximize steady-state lifetime welfare and per-worker output at the theoretical level is the research question of this paper.

\section{The Model}
%% \label{}
Consider an overlapping-generations (OLG) model of \cite{Diamond1965} in which each individual lives two periods: adulthood and old-age. The number of individuals $L$ born at time $t$ ($t=1,2,\cdots, T, T \to \infty$) is constant. As adults, individuals are endowed with one unit of labor which they supply inelastically to the market. In the model, we introduce the environmental concerns and an associated health feedback effect. The recycling of environmental tax revenues towards pollution abatement and labor income is also taken into account.

\subsection{Pollution, health and the economic structure}

Following \cite{Pautrel2012}, the pollution  stock $P$ at time $t+1$ is assumed to increase with pollution emissions ${E}_{t+1}$, decrease with the natural pollution absorption and abatement activities  ${D}_{t+1}$ funded by the government:
\begin{equation}\label{eq:r1}
{P_{t + 1}} = {\left[ {\frac{{{E_{t + 1}}}}{{{D_{t + 1}}}}} \right]^\gamma } + (1 - \mu ){P_t},\gamma  > 0,0 < \mu  < 1,
\end{equation}
where $\gamma$ is the exogenous elasticity of pollution with respect to the ratio of emissions to abatement services $E/D$, and $\mu$ is the natural  pollution absorption rate. The pollution emissions ${E}_{{t + 1}}$ arise from final production ${Y}_{{t + 1}}$ such that
\begin{equation}\label{eq:r2}
{E_{t + 1}} = z{Y_{t + 1}},0 < z < 1,
\end{equation}
where $z$ measures the polluting capacity of the technology. Abatement activities ${D}_{{t + 1}}$ are financed by a portion ($1 - \beta$)  of the environmental tax revenues on the source of pollution ${Y}_{{t + 1}}$, this is
\begin{equation}\label{eq:r3}
{D_{t + 1}} = \tau (1 - \beta ){Y_{t + 1}},0 < \tau  \le 1,0 \le \beta  < 1.
\end{equation}

Following \cite{Pautrel2009}, we assume that public health at time $t$  (denoted by $h_{t}$) is influenced negatively by the endogenous pollution stock ${P_t}$
and positively by the exogenous part of public health expenditures in GDP (denoted by $\theta$) . This is
\begin{equation}\label{eq:r4}
{h_t} = \frac{{\eta \theta }}{{\xi P_t^\varphi }},\theta>0,\eta>0,\xi>0,\varphi>0,
\end{equation}
where $\eta$  is the productivity of the health sector, $\xi$ is a positive parameter, and $\varphi$  measures the influence of pollution on public health: a higher value of $\varphi$  means that pollution harms more public health. This assumption imposes that $P>1$.

The health status of an individual has an important implication for labor market outcomes. A worse health status makes a worker less productive. To capture the simple intuition, we assume that final goods ${Y}_{t}$ are produced by using capital ${K}_{t}$  and efficiency units of labor
$h_t^\varepsilon {L_t}$  with a Cobb-Douglas technology. This is,
${Y_t} = AK_t^\alpha {(h_t^\varepsilon {L_t})^{1 - \alpha }},\alpha  \in \left( {0,1} \right)$, where $A>0$  is a constant productivity parameter, and $\varepsilon \ge 0$  captures the effect of health on labor productivity. Output per worker is then
\begin{equation}\label{eq:r5}
y_t = Ak_t^\alpha {(h_t^\varepsilon )^{1 - \alpha }}
\end{equation}
with $k$  denoting capital per worker. Under perfect competition, firms maximize their profit ${\pi _t} = (1 - \tau ){y_t} - {R_t}{k_t} - {w_t}$. The wage per effective unit of labor $w_{t}$ and interest factor ${R_t}$ are given, respectively, by
\begin{equation}\label{eq:r10}
{w_t} = A(1 - \tau )(1 - \alpha )k_t^\alpha {(h_t^\varepsilon )^{1 - \alpha }},
\end{equation}
\begin{equation}\label{eq:r11}
{R_t} = A(1 - \tau )\alpha k_t^{\alpha  - 1}{(h_t^\varepsilon )^{1 - \alpha }}.
\end{equation}

Consider a consumer who is born at time $t$. Let $c_{1t}$ and $c_{2t+1}$ denote her consumption when young and old, respectively. Her utility function is
\begin{equation}\label{eq:r6}
{U_t} = \ln {c_{1t}} + \frac{{\ln {c_{2t + 1}}}}{{1 + \rho }},\rho  > 0.
\end{equation}
Here $\rho $ denotes the subjective discount rate. The consumer working at $t$ earns a wage income $w_{t}$ and a lump-sum transfer from environmental tax $\beta\tau y_{t}$. She consumes an amount $c_{1t}$, and saves the remainder of her revenue $s_{t}$. So the budgets constraints are
\begin{equation}\label{eq:r7}
{c_{1t}} + {s_t} = {w_t} + \beta \tau {y_t},
\end{equation}
\begin{equation}\label{eq:r8}
{c_{2t + 1}} = {R_t}{s_t}.
\end{equation}
The first-order condition for this utility maximization problem is given by
\begin{equation}\label{eq:r9}
{s_t} = {\delta}\left( {{w_t} + \beta \tau {y_t}} \right),
\end{equation}
where the savings propensity ${\delta} = \frac{1}{{2 + {\rho}}}$ is a decreasing function of the subjective discount rate.

\subsection{General equilibrium}
In this model, we assume full depreciation on the capital. Therefore, competitive equilibrium can be defined by\footnote{Note that ${k_{t + 1}} = {s_t} = \delta \left( {{w_t} + \beta \tau {y_t}} \right) = \delta \left[ {\left( {1 - \alpha } \right)\left( {1 - \tau } \right) + \beta \tau } \right]{y_t}$.}
\begin{equation}\label{eq:r12}
{k_{t + 1}} = \delta \left[ {\left( {1 - \alpha } \right)\left( {1 - \tau } \right) + \beta \tau } \right]{y_t}
\end{equation}

Now it allows us to characterize the steady state as a general equilibrium. In other words, per-worker capital, individual health status, pollution stock, per-worker output, and the wage rate remain constant at any time at ${k^*}$, ${h^*}$, ${P^*}$, ${y^*}$ and ${w^*}$, respectively. From Eqs.\eqref{eq:r1}-\eqref{eq:r4}, we have
\begin{equation}\label{eq:r18}
{P^*} = P(\beta) \equiv {\left( {z/(1 - \beta )\tau } \right)^\gamma }/\mu,
 \end{equation}
\begin{equation}\label{eq:r13}
{h^*} = {\rm{H}}(\beta ) \equiv \frac{{\eta \theta {\mu ^\varphi }}}{\xi }{\left[ {\frac{{(1 - \beta )\tau }}{z}} \right]^{\varphi \gamma }}.
\end{equation}
Consequently, health status is positively affected by the pollution abatement activities. From Eqs.\eqref{eq:r5} and \eqref{eq:r12}, the steady-state value of the per-worker capital stock is
${k^*} = k\left( \beta  \right) \equiv {\left\{ {\delta \left[ {\left( {1 - \alpha } \right)\left( {1 - \tau } \right) + \beta \tau } \right]} \right\}^{\frac{1}{{1 - \alpha }}}}{({h^*})^\varepsilon }$.
The steady-state value of per-worker output as a function of environmental tax swaps is:
\begin{equation}\label{eq:r16}
{y^*} = y\left( \beta  \right) \equiv \Phi {\left[ {\left( {1 - \alpha } \right)\left( {1 - \tau } \right) + \beta \tau } \right]^{\frac{\alpha }{{1 - \alpha }}}}{\left( {1 - \beta } \right)^{\varphi \gamma \varepsilon }}
\end{equation}
with $\Phi  = {A^{\frac{1}{{1 - \alpha }}}}{\delta ^{\frac{\alpha }{{1 - \alpha }}}}{\left( {\frac{{\eta \theta {\mu ^\varphi }}}{\xi }} \right)^\varepsilon }{\left( {\frac{\tau }{z}} \right)^{\varphi \gamma \varepsilon }}$. The steady-state wage rate is given by
\begin{equation}\label{eq:r14}
{w^*} = w(\beta ) \equiv \Phi \left( {1 - \alpha } \right)\left( {1 - \tau } \right){\left[ {\left( {1 - \alpha } \right)\left( {1 - \tau } \right) + \beta \tau } \right]^{\frac{\alpha }{{1 - \alpha }}}}{\left( {1 - \beta } \right)^{\varphi \gamma \varepsilon }}
\end{equation}

Finally, using Eqs.\eqref{eq:r6}-\eqref{eq:r12} and \eqref{eq:r16}, we obtain the lifetime welfare along the balanced growth path as follows:
\begin{equation}\label{eq:r20}
{U^*} = \ln \left( {1 - \delta } \right){\Phi ^{\frac{1}{{1 - \delta }}}}{\left[ {\alpha \left( {1 - \tau } \right)} \right]^{\frac{\delta }{{1 - \delta }}}} + \ln \Omega \left( \beta  \right)
\end{equation}
with $\Omega \left( \beta  \right) = {\left[ {\left( {1 - \alpha } \right)\left( {1 - \tau } \right) + \beta \tau } \right]^{\frac{\alpha }{{\left( {1 - \alpha } \right)\left( {1 - \delta } \right)}} + 1}}{\left( {1 - \beta } \right)^{\frac{{\varphi \gamma \varepsilon }}{{1 - \delta }}}}$.

\bigskip
\emph{Proposition 1. (i) Below (respectively above) a share of environmental tax revenues towards labor income ($\hat \beta$) defined as
\begin{equation}\label{eq:r19}
\hat \beta =\frac{{\frac{\alpha }{{1 - \alpha }} - \varphi \gamma \varepsilon \left( {1 - \alpha } \right)\left( {\frac{1}{\tau } - 1} \right)}}{{\frac{\alpha }{{1 - \alpha }} + \varphi \gamma \varepsilon }}
\end{equation}
if $\frac{{\varphi r\varepsilon {{(1 - \alpha )}^2}}}{{\alpha  + \varphi r\varepsilon {{(1 - \alpha )}^2}}} < \tau  \le 1$, a higher shift to labor income raises (respectively lowers) the steady-state level of output per worker ${y^*}$. \\
(ii) $\hat \beta = 0$, if $0 < \tau  \le \frac{{\varphi r\varepsilon {{(1 - \alpha )}^2}}}{{\alpha  + \varphi r\varepsilon {{(1 - \alpha )}^2}}}$}.

\bigskip
\emph{Proof.} The influence of recycling environmental tax revenues on the steady-state output per worker is given by\\
$\partial {y^*}/\partial \beta  = {y^*}\left\{ {\frac{\alpha \tau}{{(1 - \alpha )\left[ {(1 - \alpha )(1 - \tau ) + \beta \tau } \right]}} - \frac{{\varphi \gamma \varepsilon }}{{1 - \beta }}} \right\}$.\\
The sign of $\partial {y^*}/\partial \beta$ is positive if
$\frac{{\alpha \tau }}{{(1 - \alpha )\left[ {(1 - \alpha )(1 - \tau ) + \beta \tau } \right]}} - \frac{{\varphi \gamma \varepsilon }}{{1 - \beta }} > 0$.\\
The LHS of the inequality is a decreasing monotonic function of $\beta  \in [0,1)$ with $\mathop {\lim }\limits_{\beta  \to 1}  =  - \infty $ and $\mathop {\lim }\limits_{\beta  \to 0}  = \alpha \tau /\left[ {{{(1 - \alpha )}^2}(1 - \tau )} \right] - \varphi \gamma \varepsilon $. Thus, there is a unique $\hat \beta$ under which the inequality is verified. If $\mathop {\lim }\limits_{\beta  \to 0}  > 0$, i.e., $\varphi r\varepsilon {(1 - \alpha )^2}/\left[ {\alpha  + \varphi r\varepsilon {{(1 - \alpha )}^2}} \right] < \tau  \le 1$, the solution for $\widehat\beta $ is defined by $\alpha \tau /\left\{ {(1 - \alpha )\left[ {(1 - \alpha )(1 - \tau ) + \beta \tau } \right]} \right\} = \varphi \gamma \varepsilon /\left( {1 - \beta } \right)$. This is,
 $\hat \beta  = \left\{ {\left[ {\alpha /\left( {1 - \alpha } \right)} \right] - \varphi \gamma \varepsilon \left( {1 - \alpha } \right)\left( {{\tau ^{ - 1}} - 1} \right)} \right\}/\left\{ {\left[ {\alpha /\left( {1 - \alpha } \right)} \right] + \varphi \gamma \varepsilon } \right\}$. For $\beta  < \hat\beta $ and $\beta  > \hat \beta $, we have $\partial {y^*}/\partial \beta >0 $ and $\partial {y^*}/\partial \beta <0 $, respectively. If $\mathop {\lim }\limits_{\beta  \to 0} \leq 0$,
then $\hat\beta  = 0$. When $\varphi  = 0$, the LHS of the inequality is independent of ($1 - \beta$) and positive, and therefore we have $\partial {y^*}/\partial \beta  > 0$.

\bigskip
In this paper, it is a policy mix for recycling tax revenues. A larger returning towards labor income implies a lower returning towards abatement activities (and vice versa).
When the given environmental tax ($\tau$) exceeds the value of $\frac{{\varphi r\varepsilon {{(1 - \alpha )}^2}}}{{\alpha  + \varphi r\varepsilon {{(1 - \alpha )}^2}}}$, there is a need to diminish such a huge income distortion by returning tax revenues towards income. Indeed, we are interested in how the recycling of environmental tax revenues affects per-worker output.
As shown in  Eq.\eqref{eq:r16},
on the one hand, pollution abatement activities $(1-\beta)\tau$ funded by environmental tax  contribute to benefiting the health status of workers, thus  increasing labor productivity (positive, see ${\left( {1 - \beta } \right)^{\varphi \gamma \varepsilon }}$). On the other hand, the sum-lump transfer $\beta\tau$ would reduce the labor income distortion  caused by environmental tax $\left( {1 - \alpha } \right)\left( {1 - \tau } \right)$, leading to a higher level of steady-state savings and physical-capital (positive, see ${\left[ {\left( {1 - \alpha } \right)\left( {1 - \tau } \right) + \beta \tau } \right]^{\frac{\alpha }{{1 - \alpha }}}}$).

When $\varphi  = 0$, public health is independent of pollution, and labor productivity is not affected by the investment of pollution abatement, i.e., ${y^*}$
is independent of $(1-\beta)\tau$. In such a case, the positive effect of abatement activities no longer holds.
Hence, all of environmental tax revenues are supposed to be shifted towards labor income without consideration of pollution-related health damages.

\bigskip
\emph{Proposition 2. When the environmental tax is defined at $\left( {\frac{{\varphi r\varepsilon {{(1 - \alpha )}^2}}}{{\alpha  + \varphi r\varepsilon {{(1 - \alpha )}^2}}},1} \right]$, the returning of tax revenues towards labor income will be more likely to improve the steady state level of output per worker if the share of labor in output ($1-\alpha$) is low, the environmental tax $\tau$ is high, the influence of pollution on health $\varphi$ is low, the effect of health on labor productivity $ \varepsilon$ is low, and the elasticity of pollution with respect to the ration of emissions to abatement activities $\gamma$ is low}.

\bigskip
\emph{Proof.} From Eq.\eqref{eq:r19}, it is straightforward that $\partial \hat\beta /\partial \alpha  > 0$, $\partial \hat \beta /\partial \tau  > 0$, $\partial \hat \beta /\partial \varphi  < 0$, $\partial \hat \beta /\partial \varepsilon  < 0$, and $\partial \hat \beta /\partial \gamma  < 0$.

\bigskip
\emph{Proposition 3.
(i) Below (respectively above) the optimal recycling of environmental tax revenues towards labor income ${{\hat \beta }_U}$ defined by
\begin{equation}\label{eq:r17}
{{\hat \beta }_U} = \frac{{\frac{\alpha }{{1 - \alpha }} + 1 - \delta  - \varphi \gamma \varepsilon \left( {1 - \alpha } \right)\left( {\frac{1}{\tau } - 1} \right)}}{{\frac{\alpha }{{1 - \alpha }} + \varphi \gamma \varepsilon  + 1 - \delta }}
\end{equation}
with $\frac{{\varphi \gamma \varepsilon {{(1 - \alpha )}^2}}}{{\delta \alpha  + 1 - \delta  + \varphi \gamma \varepsilon {{(1 - \alpha )}^2}}} < \tau  \le 1$, a higher shift to labor income raises (respectively lowers)  lifetime welfare.\\
(ii) ${{\hat \beta }_U} = 0$ if  $0 < \tau  \le \frac{{\varphi \gamma \varepsilon {{(1 - \alpha )}^2}}}{{\delta \alpha  + 1 - \delta  + \varphi \gamma \varepsilon {{(1 - \alpha )}^2}}}$}.

\bigskip
\emph{Proof.} From   Eq.\eqref{eq:r20}, we can obatin\\
$\frac{{\partial {U^*}}}{{\partial \beta }} = \frac{{\tau \left( {1 + \alpha \delta  - \delta } \right)}}{{\left[ {\left( {1 - \alpha } \right)\left( {1 - \tau } \right) + \beta \tau } \right]\left( {1 - \alpha } \right)\left( {1 - \delta } \right)}} - \frac{{\varphi \gamma \varepsilon }}{{\left( {1 - \delta } \right)\left( {1 - \beta } \right)}}$.\\
Because the LHS of the inequality is a monotonic decreasing function of $\beta  \in [0,1)$ with $\mathop {\lim }\limits_{\beta  \to 1}  =  - \infty $ and $\mathop {\lim }\limits_{\beta  \to 0}  = \tau \left( {1 + \alpha \delta  - \delta } \right)/\left[ {{{\left( {1 - \alpha } \right)}^2}\left( {1 - \delta } \right)\left( {1 - \tau } \right)} \right] - \varphi \gamma \varepsilon /\left( {1 - \delta } \right)$. Hence, there is a unique ${{\hat \beta }_U}$ under which the inequality is verified. This is defined by $\partial {U^*}/\partial \beta  = 0 $ if $\mathop {\lim }\limits_{\beta  \to 0}  > 0$. In other words,
${{\hat \beta }_U} = \left[ {\frac{\alpha }{{1 - \alpha }} + 1 - \delta  - \varphi \gamma \varepsilon \left( {1 - \alpha } \right)\left( {\frac{1}{\tau } - 1} \right)} \right]/\left( {\frac{\alpha }{{1 - \alpha }} + \varphi \gamma \varepsilon  + 1 - \delta } \right)$ \\
if $\varphi r\varepsilon {(1 - \alpha )^2}/\left[ {\delta \alpha  + 1 - \delta  + \varphi r\varepsilon {{(1 - \alpha )}^2}} \right] < \tau  \le 1$. When $\beta  < {{\hat \beta }_U}$ and $\beta  > {{\hat \beta }_U}$, we have $\partial {U^*}/\partial \beta  > 0$ and $\partial {U^*}/\partial \beta  < 0$, respectively. If $\mathop {\lim }\limits_{\beta  \to 0}  \le 0$, then ${{\hat \beta }_U} = 0$.

\bigskip
Propositions 1 and 3 lead to the following corollary.

\bigskip
\emph{Corollary 1. (i) If $0 < \tau  \le \frac{{\varphi r\varepsilon {{(1 - \alpha )}^2}}}{{\delta \alpha  + 1 - \delta  + \varphi r\varepsilon {{(1 - \alpha )}^2}}}$,
$\hat \beta  = {{\hat \beta }_U} = 0$. When  $0 < \beta  < 1$, then   $\partial {y^*}/\partial \beta  < 0$ and $\partial {U^*}/\partial \beta  < 0$.\\
(ii) If $\frac{{\varphi r\varepsilon {{(1 - \alpha )}^2}}}{{\delta \alpha  + 1 - \delta  + \varphi r\varepsilon {{(1 - \alpha )}^2}}} < \tau  \le \frac{{\varphi r\varepsilon {{(1 - \alpha )}^2}}}{{\alpha  + \varphi r\varepsilon {{(1 - \alpha )}^2}}}$, then ${{\hat \beta }_U} >  \hat \beta = 0$. When
$0 < \beta  < {{\hat \beta }_U}$ then $\partial {y^*}/\partial \beta  < 0$ and $\partial {U^*}/\partial \beta  > 0$, and when ${\hat \beta _U} < \beta  < 1$
 then $\partial {y^*}/\partial \beta  < 0$ and $\partial {U^*}/\partial \beta  < 0$.\\
(iii) If $\frac{{\varphi r\varepsilon {{(1 - \alpha )}^2}}}{{\alpha  + \varphi r\varepsilon {{(1 - \alpha )}^2}}} < \tau  \le 1$, then ${{\hat \beta }_U} > \hat \beta  > 0$. When
$0 <\beta  < \hat \beta $ then $\partial {y^*}/\partial \beta  > 0$ and $\partial {U^*}/\partial \beta  > 0$. When $\widehat\beta  < \beta  < {{\hat \beta }_U}$ then $\partial {y^*}/\partial \beta  < 0$ and $\partial {U^*}/\partial \beta  > 0$,  and when ${\hat \beta _U} < \beta  < 1$ then $\partial {y^*}/\partial \beta  < 0$ and $\partial {U^*}/\partial \beta  < 0$}.\\

\bigskip
\emph{Proof.} From Eqs.\eqref{eq:r19}-\eqref{eq:r17}, we can obtain\\
${{\hat \beta }_U} - \hat \beta  = \frac{{\varphi \gamma \varepsilon \left( {1 - \delta } \right)\left[ {\tau  + \left( {1 - \alpha } \right)\left( {1 - \tau } \right)} \right]}}{{\tau \left( {\frac{\alpha }{{1 - \alpha }} + \varphi \gamma \varepsilon  + 1 - \delta } \right)\left( {\frac{\alpha }{{1 - \alpha }} + \varphi \gamma \varepsilon } \right)}} > 0$
if $\frac{{\varphi r\varepsilon {{(1 - \alpha )}^2}}}{{\alpha  + \varphi r\varepsilon {{(1 - \alpha )}^2}}} < \tau  \le 1$.

\bigskip
Corollary 1 means that the positive impact of a larger shift towards labor income on the steady-state lifetime welfare might have a detrimental impact on per-worker output if the environmental tax $\tau$ is defined from $\frac{{\varphi r\varepsilon {{(1 - \alpha )}^2}}}{{\delta \alpha  + 1 - \delta  + \varphi r\varepsilon {{(1 - \alpha )}^2}}}$  to 1.
This follows from the fact that the net welfare effect is the sum of an output effect and an income effect (see the last term in Eq. \eqref{eq:r20}). As a result, if $\frac{{\varphi \gamma \varepsilon {{(1 - \alpha )}^2}}}{{\delta \alpha  + 1 - \delta  + \varphi \gamma \varepsilon {{(1 - \alpha )}^2}}} < \tau  \le 1$,
when $ 0 <\beta  < \hat \beta $ a larger shift towards labor income leads to a positive output effect and a positive income effect, thus enhancing  both the steady-state lifetime welfare and per-worker output; when $\widehat\beta  < \beta  < {{\hat \beta }_U}$  a larger shift leads to a negative output effect but a positive income effect, and the former effect is smaller than the latter, thus enhancing the lifetime welfare but reducing per-worker output; when ${{\hat \beta }_U}< \beta  < 1$  a larger shift leads to a negative output effect and a positive income effect, and the former effect exceeds the latter, thus diminishing both the steady-state lifetime welfare and per-worker output.

\section{Conclusions}
This paper takes the pollution-related health damage into account in a simple OLG model and finds that the recycling of environmental tax revenues towards pollution abatement and labor income contributes to maximizing steady-state economy variables (per-worker output and welfare). One policy implication is the need to continue and reinforce efforts to provide a substantial shift of tax revenues towards abatement activity and a relatively small shift towards labor income in economies where the health is very pollution-sensitive and where industries is labor-intensive.
The other one is that the per-worker output implication of any environmental policy should be carefully taken into account to avoid the situation where there is a negative impact on per-worker output from the allocation of environmental tax revenues while lifetime welfare is increased.

%\section*{Acknowledgements}
%We gratefully acknowledge the financial support from the Program for New Century Excellent Talents in University of the Minisrty of Education of China (Grant Nos. NCET-10-0779
%and 2013RC020) and the National Natural Science Foundation of China (Grant Nos. 71001101 and 71273261).

\bibliography{myarticle}
\bibliographystyle{econ_bulletin}

\end{document}